\documentclass[12pt,english]{article}
\pdfoutput=1
\usepackage[T1]{fontenc}
\usepackage{graphicx,array}
\usepackage{cite}
\usepackage{color}
\usepackage{latexsym}
\usepackage{amsthm}
\usepackage{amsmath}
\usepackage[titletoc]{appendix}
\usepackage{enumitem}
\usepackage{amssymb}
\usepackage{color}
\usepackage[unicode=true,
 bookmarks=true,bookmarksnumbered=false,bookmarksopen=false,
 breaklinks=false,pdfborder={0 0 1},backref=false,colorlinks=true]
 {hyperref}
\hypersetup{pdftitle={New Spacetimes for Rotating Dust in (2+1)-Dimensional General Relativity},
 pdfauthor={Grant N. Remmen},
 citecolor=black,linkcolor=black,urlcolor=black}
\usepackage{breakurl}
\usepackage[hang,flushmargin]{footmisc} 

\def\simgt{\mathrel{\lower2.5pt\vbox{\lineskip=0pt\baselineskip=0pt
           \hbox{$>$}\hbox{$\sim$}}}}
\def\simlt{\mathrel{\lower2.5pt\vbox{\lineskip=0pt\baselineskip=0pt
           \hbox{$<$}\hbox{$\sim$}}}}

\newcommand{\be}{\begin{equation}}
\newcommand{\ee}{\end{equation}}
\newcommand{\bea}{\begin{eqnarray}}
\newcommand{\eea}{\end{eqnarray}}
\newcommand{\Eq}[1]{Eq.~(\ref{#1})}
\newcommand{\Eqs}[2]{Eqs.~(\ref{#1}) and (\ref{#2})}
\newcommand{\Sec}[1]{Sec.~\ref{#1}}

\newcommand{\Fig}[1]{Fig.~\ref{#1}}
\newcommand{\App}[1]{App.~\ref{#1}}

\newcommand{\Ref}[1]{Ref.~\cite{#1}}

\hypersetup{
 colorlinks,citecolor=black,linkcolor=black,urlcolor=black}
\usepackage{xcolor}

\begin{document}

\interfootnotelinepenalty=10000
\baselineskip=18pt
\hfill

\vspace{2cm}
\thispagestyle{empty}
\begin{center}
{\LARGE \bf
New Spacetimes for Rotating Dust in $(2+1)$-Dimensional General Relativity
}\\
\bigskip\vspace{1cm}{
{\large Grant N. Remmen}
} \\[7mm]
{\it Center for Theoretical Physics and Department of Physics \\
     University of California, Berkeley, CA 94720, USA and \\
     Lawrence Berkeley National Laboratory, Berkeley, CA 94720, USA}
\let\thefootnote\relax\footnote{e-mail: 
\url{grant.remmen@berkeley.edu}}
 \end{center}

\bigskip
\centerline{\large\bf Abstract}
\begin{quote} \small
Multi-parameter solutions to the Einstein equations in $2+1$ dimensions are presented, with stress-energy given by a rotating dust with negative cosmological constant. The matter density is uniform in the corotating frame, and the ratio of the density to the vacuum energy may be freely chosen. The rotation profile of the dust is controlled by two parameters, and the circumference of a circle of a given radius is controlled by two additional parameters. Though locally related to known metrics, the global properties of this class of spacetimes are nontrivial and allow for new and interesting structure, including apparent horizons and closed timelike curves, which can be censored by a certain parameter choice. General members of this class of metrics have two Killing vectors, but parameters can be chosen to enhance the symmetry to four Killing vectors. The causal structure of these geometries, interesting limits, and relationship to the G\"odel metric are discussed. An additional solution, with nonuniform dust density in a Gaussian profile and zero cosmological constant, is also presented, and its relation to the uniform-density solutions in a certain limit is discussed.

\end{quote}
	
\setcounter{footnote}{0}

\newpage
\tableofcontents
\newpage

\section{Introduction}
\label{sec:Introduction}

In the past century of Einstein's theory of gravity, progress in general relativity has been shaped in large part by the discovery of new exact solutions to the field equations.
From Schwarzschild's work to the present day, exact solutions provide useful laboratories for investigating the properties of relativistic gravitating systems.

Two exact solutions of particular interest are G\"odel's spacetime \cite{Godel} describing a rotating dust and Ba\~nados, Teitelboim, and Zanelli's black hole \cite{BTZ}. Both of these solutions describe geometries in $2+1$ dimensions, with negative cosmological constant. G\"odel's solution demonstrated properties of closed timelike curves (CTCs) in general relativity, while the BTZ geometry showed that black holes can exist in three-dimensional gravity. General relativity in three dimensions provides an interesting arena in which to explore the geometric properties of spacetime, without the complications endemic to a nonzero Weyl tensor \cite{Brill:1998pr}. Moreover, $(2+1)$-dimensional spacetimes with negative cosmological constant are of current interest as a result of the AdS/CFT correspondence \cite{Maldacena:1997re,Gubser:1998bc,Witten:1998qj,Aharony:1999ti}, in particular, aspects of the AdS$_3$/CFT$_2$ case (see \Ref{Kraus:2006wn} and refs. therein for a review, as well as \Ref{Brown:1986nw}).

In this paper, a class of axially-symmetric spacetime geometries in $2+1$ dimensions is presented. The cosmological constant is negative, and the energy-momentum tensor is described by pressureless, rotating dust, as in the G\"odel solution. However, this class of geometries is much more general, with multiple free parameters describing the dust density, (spatially-dependent) rotation rate, and circumference profile of the spacetime. Depending on the parameters, these spacetimes can exhibit many interesting properties, including CTCs, apparent horizons, spacetime boundaries, enhanced symmetry algebras, and geodesic completeness.

This paper is organized as follows. 
In \Sec{sec:metric}, the class of solutions is stated and the corotating timelike congruence is examined. Next, in \Sec{sec:causal} the causal structure of the spacetime is investigated, including null congruences, Cauchy horizons, geodesic completeness, and censorship of the CTCs. 
Particular limiting geometries are discussed in \Sec{sec:limits} and the symmetries of the spacetime are found in \Sec{sec:isometries}.
We examine various special cases of interest in \Sec{sec:special} and conclude with the discussion in \Sec{sec:discussion}.
In \App{app:Gaussian}, another new spacetime geometry is exhibited, with a spinning dust with Gaussian density profile, which in a certain limit reduces to a member of the family of geometries we present in \Sec{sec:metric}.

\section{A Class of Spinning Dust Solutions}\label{sec:metric}

Consider the following stationary metric in 2+1 dimensions:
\be
{\rm d}s^{2}=-{\rm d}t^{2}-w(r)\,{\rm d}t\,{\rm d}\phi+D(r)\,{\rm d}\phi^{2}+\frac{{\rm d}r^{2}}{N(r)},\label{eq:metric}
\ee
where
\be
\begin{aligned}w(r) & =2L\left[j_{1}\left(\frac{L}{r}\right)+j_{2}\right]\\
D(r) & =-j_{1}^{2}ML^{2}\left(\frac{L}{r}\right)^{2}+2c_{1}L^{2}\left(\frac{L}{r}\right)+c_{2}L^{2}\\
N(r) & =4\left(1-M\right)\left(\frac{r}{L}\right)^{2}+8\left(\frac{j_{1}j_{2}+c_{1}}{j_{1}^{2}}\right)\left(\frac{r}{L}\right)^{3}+4\left(\frac{j_{2}^{2}+c_{2}}{j_{1}^{2}}\right)\left(\frac{r}{L}\right)^{4}.
\end{aligned}
\label{eq:funs}
\ee
Here, $j_{1}$, $j_{2}$, $c_{1}$, $c_{2}$, and $M$ are unitless
constants and $-\infty<t<\infty$, $0\leq r<\infty$, and $0\leq\phi<2\pi$.\footnote{As we will show in \Sec{sec:discussion}, having chosen a value of $L$, which specifies the overall length scale, the coefficients $(M,j_1,j_2,c_1,c_2)$ indeed specify a five-parameter family of distinct geometries; that is, for generic values of these constants, the geometry one obtains is not equivalent to another member of this family under diffeomorphism.}
This metric describes a rotating, pressureless dust solution with
density $4M/L^{2}$ and with negative cosmological constant $\Lambda=-1/L^{2}$.
Defining $u^{a}=\partial_{t}=(1,0,0)$ for the reference frame of the dust,\footnote{Unless indicated otherwise, we use an axial $(t,r,\phi)$ coordinate system throughout.}
the Einstein equations are satisfied:
\be
R_{ab}-\frac{1}{2}R\,g_{ab} + \Lambda \,g_{ab}=\frac{4M}{L^{2}}u_{a}u_{b}.
\ee
The metric \eqref{eq:metric} has several interesting properties.
Frame-dragging is evidenced by the nonzero ${\rm d}t\,{\rm d}\phi$ component of the metric.
However, the dust in this solution does not rotate at constant rate,
but instead at a rate dictated by $w(r)$. In particular, if $j_{1}$
and $j_{2}$ have opposite sign, then the dust rotates clockwise for
small $r$ and counterclockwise for large $r$, or vice versa, and has vanishing rotation at some radius $r_0 = -\frac{j_1}{j_2}L$. We
note that $w(r)$, $D(r)$, and $N(r)$ in \Eq{eq:funs} are related by the useful identity
\be
j_{1}^{2}L^{6}N(r)=r^{4}\left[4D(r)+w^{2}(r)\right].\label{eq:useful}
\ee

Let us consider a timelike geodesic congruence corotating with the
dust. We choose the initial tangent vector for our congruence of geodesics
to be $u^{a}$, and since
\be
u^{a}\nabla_{a}u^{b}=0,
\ee
we obtain a family of timelike geodesics that continue to point along
$u^{a}$ at all times.
Since we can show that $u^{a}$ is a global timelike Killing vector of the geometry
\eqref{eq:metric}, the spacetime is stationary. However, it is not static
for $j_{1}\neq0$, since there do not exist hypersurfaces everywhere
orthogonal to orbits of the timelike Killing vectors, or equivalently
by Frobenius's theorem \cite{Wald}, $u_{[a}\nabla_{\vphantom{]}b}u_{c]}\neq0$.

For this timelike congruence, we can define a spatial metric $h_{ab}=g_{ab}+u_{a}u_{b}$; 
the extrinsic curvature is $B_{ab}=\nabla_{b}u_{a}$. Then we
find that the expansion, measuring the logarithmic derivative of the
area element along the congruence, vanishes:
\be
\theta=h^{ab}B_{ab}=0.
\ee
Similarly, the shear vanishes, implying that the dust is rigidly rotating:\footnote{A theorem of G\"odel \cite{Godel,Ellis} states that if a spacetime possesses spacelike homogeneous hypersurfaces and a timelike geodesic congruence with vanishing expansion and shear, then the metric must locally correspond to either the G\"odel universe or the Einstein static universe. For the metric \Eq{eq:metric}, the matter density is homogeneous, $M = \text{constant}$, but the metric is not. Hence, \Eq{eq:metric} is allowed to differ from the G\"odel or Einstein static universe, which it manifestly does in general, since $M$ is independent of $\Lambda$.}
\be
\varsigma_{ab}=B_{(ab)}-\frac{1}{2}\theta h_{ab}=0.
\ee
However, the vorticity tensor\footnote{This object is sometimes called the twist \cite{Wald}. However, to avoid confusion with the twist one-form we will later define for null congruences, we use this alternative nomenclature.} $\Omega_{ab} = B_{[ab]}$ does not vanish:
\be
\Omega_{ab} \,{\rm d}x^a \wedge {\rm d}x^b = 2\, \Omega_{r\phi} \,{\rm d}r\wedge {\rm d}\phi = -\frac{j_1 L^2}{r^2} {\rm d}r\wedge {\rm d}\phi .\label{eq:twisttensor}
\ee
Since the vorticity is nonzero, the geodesic congruence formed by $u^{a}$ is not hypersurface orthogonal \cite{Wald}. This is another manifestation of the fact
that the spacetime, while stationary, is not static.

An extensive review of other dust solutions can be found in \Ref{Krasinski}. In more recent work, Refs.~\cite{Barrow:2006cw,Podolsky:2018zha} gave the general $(2+1)$-dimensional solutions for an irrotational dust and a null dust, respectively.\footnote{In contrast, the dust solution described in this paper is timelike and has nonvanishing vorticity.} As we will see in \Sec{sec:Godel}, while our family of metrics contains the G\"odel universe as a special case, it is in general distinct from G\"odel's solution. Our metric is axially symmetric, but is not locally rotationally symmetric in the sense of \Ref{Ellis}. It is distinct from the van~Stockum \cite{VanStockum} (or more generally, Lanczos \cite{Lanczos}) solution, which has nonhomogeneous dust density.
The traceless Ricci tensor $R_{ab} - \frac{1}{3} R\,g_{ab}$ for the metric given in Eqs.~\eqref{eq:metric} and \eqref{eq:funs} is proportional to $g_{ab} + 3 u_a u_b$, so it is of Petrov-Segre type D\textsubscript{t} in the notation of \Ref{Chow:2009km}.
The metric is related by a local diffeomorphism to a timelike-squashed AdS\textsubscript{3} solution of topologically massive gravity \cite{Chow:2009km,Gurses:1994bjn} and by a different local diffeomorphism to the four-parameter family of solutions of Lubo, Rooman, and Spindel \cite{Lubo:1998ue}. 
Our class of metrics possesses five unitless parameters, plus a length scale $L$, and as we will discuss in \Sec{sec:discussion}, the family of globally distinct solutions is five-dimensional; the extra freedom comes from the fact that our metric \eqref{eq:metric} is an analytic extension of the metric in \Ref{Lubo:1998ue} and, for various parameter choices, can have different global properties.
As we will see in this work, these global properties lead to interesting physical differences in three-dimensional gravity, including considerations of topology, causal structure, horizons, boundaries, geodesic completeness, and singularities.

\section{Causal Structure}\label{sec:causal}

To investigate the causal structure of the spacetime, it is useful to first construct some null congruences. We will find apparent horizons, spacetime boundaries, and CTCs, and also discover how to censor them.

\subsection{Null congruences}\label{sec:apparent}

\begin{figure}[t]
\begin{center}
\hspace{-5mm} \includegraphics[width=\textwidth]{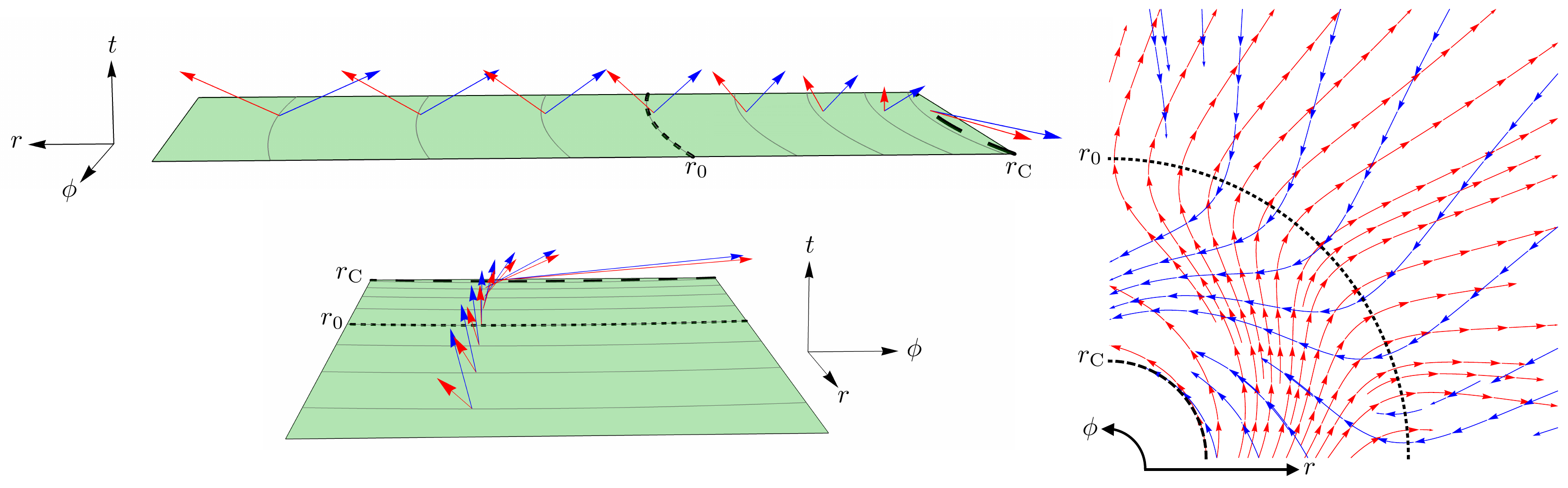}
\end{center}
\vspace{-5mm}
\caption{Illustration of the null congruences orthogonal to circles of fixed $r$ (gray arcs) on some slice of constant $t$ (green surface). The outgoing and ingoing orthogonal null congruences have tangents given in \Eq{eq:vecs} by $k^a$ and $\ell^a$ and are illustrated by red and blue arrows, respectively. For this illustration, $j_1 j_2$ was chosen to be negative, so the frame-dragging vanishes on the surface at radius $r_0$ (dotted line) where $w(r_0) = 0$ and goes in the $+\phi$ or $-\phi$ direction inside and outside, respectively. As the Cauchy horizon $r_{\rm C}$ (dashed line) is approached, the orthogonal null congruences are dragged around the angular direction. For $r<r_{\rm C}$, the congruence is not defined.}
\label{fig:congruences}
\end{figure}

Consider a surface $\sigma$ at constant $r$, for some $r$ chosen
such that $N(r)$ and $D(r)$ are positive. There are two future-pointing
null geodesic congruences generated by the null vectors with initial
tangents $k^{a}$ and $\ell^{a}$ orthogonal to $\sigma$:
\be
\begin{aligned}k^{a} & =\frac{1}{\sqrt{2}}\left(\frac{1}{\sqrt{1+\frac{w^{2}(r)}{4D(r)}}},\sqrt{N(r)},\frac{w(r)}{2D(r)\sqrt{1+\frac{w^{2}(r)}{4D(r)}}}\right)\\
\ell^{a} & =\frac{1}{\sqrt{2}}\left(\frac{1}{\sqrt{1+\frac{w^{2}(r)}{4D(r)}}},-\sqrt{N(r)},\frac{w(r)}{2D(r)\sqrt{1+\frac{w^{2}(r)}{4D(r)}}}\right).
\end{aligned}\label{eq:vecs}
\ee
Despite the fact that $k^{\phi},\ell^{\phi}\neq0$, one can verify
that $k_{a}\phi^{a}=\ell_{a}\phi^{a}=0$, for $\phi^{a}= \partial_\phi = (0,0,1)$
giving the unit tangent to $\sigma$. We note that $k^{2}=\ell^{2}=0$, and we
have chosen the relative normalization such that $k\cdot\ell=-1$.
Both $k$ and $\ell$ are future pointing (i.e., $k^{t}\geq0$ and
$\ell^{t}\geq0$), and $k$ is outward-pointing while $\ell$
is inward-pointing (i.e., $k^{r}\geq0$ and $\ell^{r}\leq0$). In \Eq{eq:vecs},
we need $1+w^{2}(r)/4D(r)\geq0$, which is justified by \Eq{eq:useful}
along with our choice of $r$ such that $N(r)$ and $D(r)$ are both positive.
See \Fig{fig:congruences} for an illustration of the congruences.

From the congruences, we define the induced metric $q_{ab}=g_{ab}+k_{a}\ell_{b}+k_{b}\ell_{a}$,
the null extrinsic curvature $B_{ab}^{(k)}=q_{a}^{\;\;c}q_{b}^{\;\;d}\nabla_{d}k_{c}$
(and analogously for $\ell$), and the null expansion $\theta_{k}[\sigma]=q^{ab}B_{ab}^{(k)}$
(which measures the logarithmic derivative of the area element along
the affine parameter of the null geodesic). After explicit computation,
we have:
\be
\theta_{k}[\sigma]=-\theta_{\ell}[\sigma]=\frac{D'(r)\sqrt{N(r)}}{2\sqrt{2}D(r)}.\label{eq:thetas}
\ee
The shear of the congruences vanishes:
\be
\begin{aligned}\varsigma_{ab}^{(k)} & =B_{(ab)}^{(k)}-\theta_{k}q_{ab}=0\\
\varsigma_{ab}^{(\ell)} & =B_{(ab)}^{(\ell)}-\theta_{\ell}q_{ab}=0,
\end{aligned}
\ee
which is a consequence of the fact that the shear tensor is by definition
traceless and that the null congruence is codimension-two (and hence
one-dimensional in this three-dimensional spacetime). The vorticity tensors
also vanish,
\be
\begin{aligned}\Omega_{ab}^{(k)} & =B_{[ab]}^{(k)}=0\\
\Omega_{ab}^{(\ell)} & =B_{[ab]}^{(\ell)}=0,
\end{aligned}
\ee
which is a consequence of the fact that the congruences are hypersurface
orthogonal.
The twist one-form gauge field (i.e., the H{\'a}{\'\j}i{\v c}ek one-form)
is
\be
\omega_{a}=\frac{1}{2}q_{ab}{\cal L}_{k}\ell^{b}=-q_{a}^{\;\;b}\ell^{c}\nabla_{b}k_{c}=\sqrt{\frac{N(r)}{1+\frac{w^{2}(r)}{4D(r)}}}\frac{w(r)D'(r)-D(r)w'(r)}{8D(r)^{2}}\times\left(w(r),0,-2D(r)\right),
\ee
where ${\cal L}_{k}$ denotes the Lie derivative along $k$.

Were we to choose parameters for which we can have $D(r)>0$ when $N(r) = 0$, the metric in \Eq{eq:metric} would have apparent horizons, where $\theta_{k}$
or $\theta_{\ell}$ vanish, at the zeros of $N(r)$ located at $r=r_{\pm}$:
\be
r_{\pm}=L\left(\frac{j_{1}j_{2}+c_{1}}{j_{2}^{2}+c_{2}}\right)\left[-1\pm\sqrt{1-\left(1-M\right)\frac{j_{1}^{2}\left(j_{2}^{2}+c_{2}\right)}{\left(j_{1}j_{2}+c_{1}\right)^{2}}}\right].\label{eq:horizons}
\ee
Whether or not these zeros exist (i.e., whether $r_{\pm}$ is real
and positive) in the spacetime depends on the relative signs and magnitudes
of $j_{1}$, $j_{2}$, $c_{1}$, $c_{2}$, and $M$. 

To guarantee that the angular direction is not timelike at large $r$, we must take $c_2 \geq 0$. If we choose $c_1 > 0$, then $D(r)$ would reach a maximum at $r = r_{\rm m} = \frac{j_1^2}{c_1}ML$. In this case, the signs of $\theta_k$ and $\theta_\ell$ would flip at $r_{\rm m}$. We would have an apparent horizon at $r_{\rm m}$, where the null expansions can vanish. 
For the rest of this paper, we will consider $c_1 \leq 0$, so that $D'(r)>0$; this ensures that circles of constant $r$ have circumferences that grow with $r$ asymptotically, as we would intuitively expect. 
In that case, $\sigma$ is a normal surface, i.e., a surface for which the outward future-pointing null congruence has positive expansion and the inward future-pointing null congruence has negative expansion. 

\subsection{Cauchy horizon and boundary}\label{sec:Cauchy}

This spacetime can exhibit Cauchy horizons, defining the boundary
of the nonchronological region of the spacetime where $D(r)<0$.
In this region, circles in $\phi$ at constant $t$ and $r$ are CTCs.\footnote{Note that the appearance of CTCs does not run afoul of the theorem in \Ref{Raeymaekers:2011dd}, since the CTCs extend arbitrarily close to $r=0$, and thus constitute ``boundary CTCs'' in the sense of \Ref{Raeymaekers:2011dd}.}
The Cauchy horizons are given by the zeros of $D(r)$:
\be
r_{\pm}^{{\rm C}}=-L\left(\frac{c_{1}}{c_{2}}\right)\left(1\pm\sqrt{1+\frac{j_{1}^{2}Mc_{2}}{c_{1}^{2}}}\right).
\ee

\begin{figure}[t]
\begin{center}
\hspace{-5mm} \includegraphics[height=0.6\textwidth]{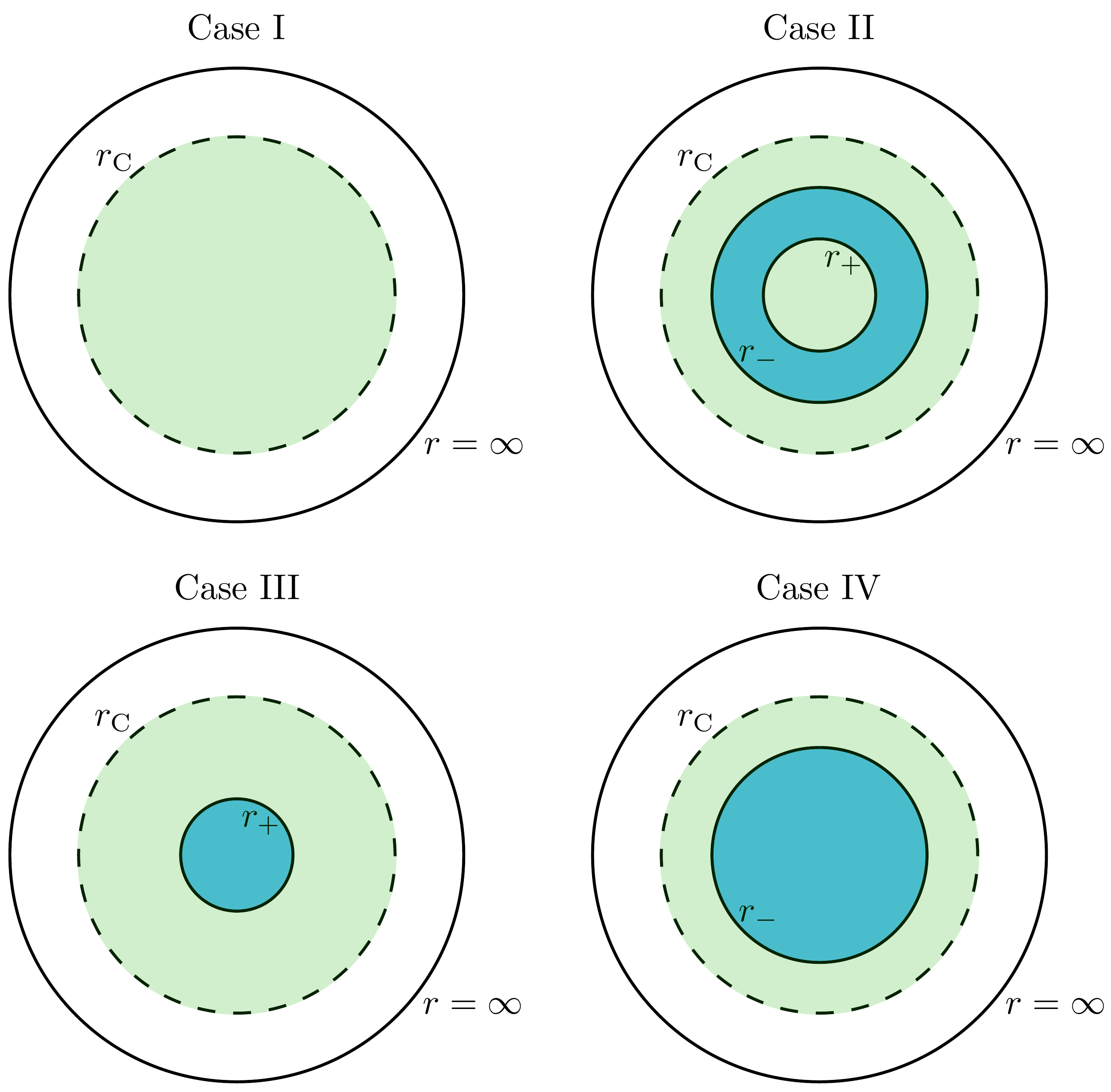}
\end{center}
\vspace{-5mm}
\caption{Cases in \Eq{eq:cases} for geometry with metric given in \Eqs{eq:metric}{eq:funs}.  The singularities in the $rr$ component of the metric (the zeros of $N(r)$) are $r_\pm$, while the Cauchy horizon $r_{\rm C}$ is determined by the zero of the $\phi\phi$ component of the metric $D(r)$. In Cases I and II, $0<M<1$, while in Cases III and IV, $M\geq 1$. In Cases I and III, $j_1 j_2 + c_1 >0$, while in Cases II and IV, $j_1 j_2 + c_1 < 0$. If $c_2 \geq 0$ and $c_1 \leq 0$, then $r_{\rm  C} \geq r_{\pm}$.}
\label{fig:cases}
\end{figure}

By the null energy condition, we will take $M\geq0$, so that the dust has nonnegative
density. Further, taking $c_2 \geq 0$ and $c_1 \leq 0$ as noted previously, the nonchronological region
is given by $r<r_{{\rm C}}$, where we write $r_{{\rm C}}$ for $r_{+}^{{\rm C}}$,
the single Cauchy horizon:
\be
r_{{\rm C}}=-L\left(\frac{c_{1}}{c_{2}}\right)\left(1+\sqrt{1+\frac{j_{1}^{2}Mc_{2}}{c_{1}^{2}}}\right).
\ee
We then have the following conditions for the existence of the zeros of $N(r)$ in \Eq{eq:horizons}:
\be
\begin{aligned}\text{Case I: } & 0<M<1,\;j_{1}j_{2}+c_{1}>0 & \implies & \text{no zero}\\
\text{Case II: } & 0<M<1,\;j_{1}j_{2}+c_{1}<0 & \implies & \text{two zeros at }r_{\pm},\;r_{+}<r_{-}\\
\text{Case III: } & M\geq1,\;j_{1}j_{2}+c_{1}>0 & \implies & \text{one zero at }r_{+}\\
\text{Case IV: } & M\geq1,\;j_{1}j_{2}+c_{1}<0 & \implies & \text{one zero at }r_{-}.
\end{aligned}\label{eq:cases}
\ee
See \Fig{fig:cases} for an illustration. The nonchronological region can extend outside the apparent
horizon, i.e., it is possible to have $r_{{\rm C}}>r_{\pm}$. 
Indeed, with our assumptions $c_1 \leq 0$ and $c_2 \geq 0$, one can show that the Cauchy horizon always satisfies $r_{\rm C} \geq r_{\pm}$, which implies that the congruence in \Sec{sec:apparent} cannot be extended down to $r_\pm$ (since the construction of the congruence required $D(r)>0$ by \Eq{eq:useful}). Hence,  if $c_1 \leq 0$ and $c_2 \geq 0$, the zeros of $N(r)$ are not truly apparent horizons where $\theta_k$ or $\theta_\ell$ vanish, since the congruence does not exist there.

In the region where $N(r)<0$, we can consider a surface
$\mu$ at constant $r$. Let us attempt to construct the orthogonal null
congruences from $\mu$. Writing one of the tangent vectors to such
a congruence as $\bar{k}^{a}$, we must have $\bar{k}_{\phi}=0$ for
the congruence to be orthogonal to $\mu$. Thus, we have
\be
\bar{k}^{2}=(\bar{k}_{r})^{2}N(r)-\frac{(\bar{k}_{t})^{2}}{1+\frac{w^{2}(r)}{4D(r)}}.\label{eq:nulltest}
\ee
Since we have $N(r)<0$ on $\mu$ and since \Eq{eq:funs} requires $j_1 \neq 0$, \Eq{eq:useful} implies that we must have $D(r)<0$ on $\mu$. By \Eq{eq:useful} again, we then have $1+w^{2}(r)/4D(r)>0$ on $\mu$,
which by \Eq{eq:nulltest} makes the requirement $\bar{k}^{2}=0$
impossible to satisfy for any choice of $\bar{k}_{t}$ and $\bar{k}_{r}$.
Hence, the null congruences from $\mu$ do not exist, i.e.,
the region $N(r)<0$ is of non-Lorentzian causal structure. We therefore
exclude this region from the spacetime and take a surface at $r= r_{\pm}$
to be a boundary of the geometry.

\subsection{Geodesics and completeness\label{sec:completeness}}

\begin{figure}[t]
\begin{center}
\hspace{-5mm} \includegraphics[height=0.3\textwidth]{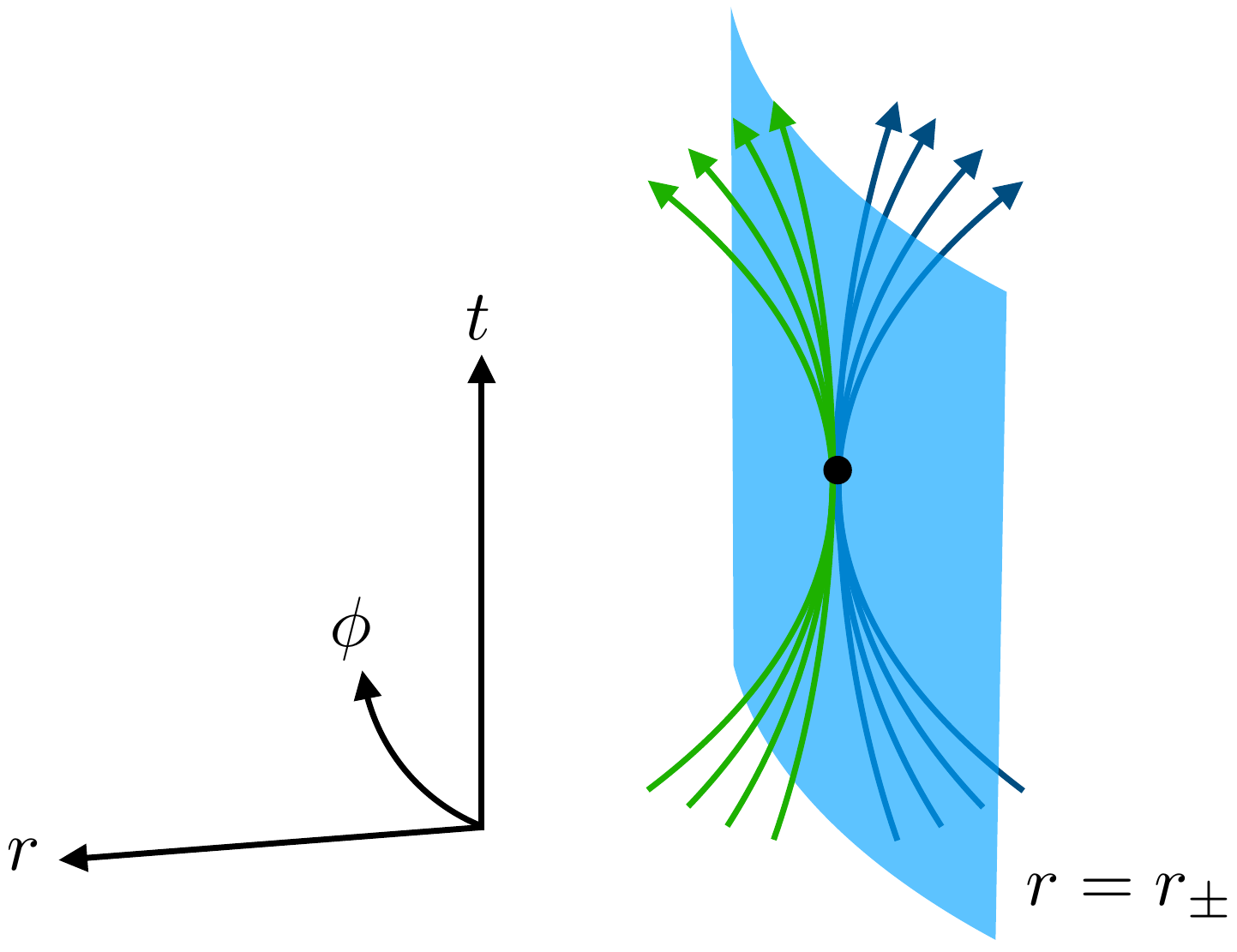}
\end{center}
\vspace{-5mm}
\caption{Geodesics bouncing off of a boundary at $r = r_\pm$. Depicted is a family of geodesics with different initial data $r_{\rm i}$ and $v_{\rm i}$ chosen to all intersect the boundary at some fixed spacetime location (black dot). The bounce occurs for geodesics originating on either side of the boundary.}
\label{fig:bounce}
\end{figure}

Let us consider the evolution of a timelike geodesic in this spacetime. The geodesic
equation,
\be
\frac{{\rm d}^{2}x^{a}}{{\rm d}\tau^{2}}+\Gamma_{bc}^{a}\frac{{\rm d}x^{b}}{{\rm d}\tau}\frac{{\rm d}x^{c}}{{\rm d}\tau}=0,
\ee
implies
\be
\begin{aligned}
\frac{j_1^2 L^6 N(r)}{r^4}\frac{{\rm d}^2 t}{{\rm d}\tau^2} -2 w(r) D'(r) \frac{{\rm d} r}{{\rm d}\tau} \frac{{\rm d}\phi}{{\rm d}\tau} + w'(r)\frac{{\rm d} r}{{\rm d}\tau} \left[2 D(r)\frac{{\rm d}\phi}{{\rm d}\tau} + w(r)\frac{{\rm d}t}{{\rm d}\tau}\right] &=0 \\
\frac{{\rm d}^{2}r}{{\rm d}\tau^{2}}-\frac{N'(r)}{2N(r)}\left(\frac{{\rm d}r}{{\rm d}\tau}\right)^{2} + \frac{N(r)}{2}\frac{{\rm d}\phi}{{\rm d}\tau}\left[w'(r)\frac{{\rm d}t}{{\rm d}\tau}-D'(r)\frac{{\rm d}\phi}{{\rm d}\tau}\right] &=0 \\
\frac{j_1^2 L^6 N(r)}{r^4}\frac{{\rm d}^2 \phi}{{\rm d}\tau^2} + \frac{{\rm d}r}{{\rm d}\tau} \left[\left(4 D'(r)+ w(r) w'(r)\right)\frac{{\rm d}\phi}{{\rm d} \tau} - 2 w'(r) \frac{{\rm d}t}{{\rm d}\tau} \right]&= 0,\end{aligned}\label{eq:geodesicfull}
\ee
where we used \Eq{eq:useful} to write $4D(r) + w^2(r)$ in terms of $N(r)$.
Now, near $r = r_\pm$, $N(r)$ is small, so to leading order in this limit, we must have 
\be
\frac{{\rm d}^{2}r}{{\rm d}\tau^{2}}-\frac{N'(r)}{2N(r)}\left(\frac{{\rm d}r}{{\rm d}\tau}\right)^{2} = 0,
\ee
where we can approximate $N(r)$ by $N'(r_\pm)(r-r_\pm)$. Thus, near $r_\pm$, we have the solution
\be
r(\tau)-r_{\pm}=r_{\rm i}-r_{\pm}+v_{\rm i}\tau+\frac{v_{\rm i}^{2}\tau^{2}}{4(r_{\rm i}-r_{\pm})},\label{eq:geodesic}
\ee
where $r_{\rm i}$ and $v_{\rm i}$ are constants. Note that $r_{\rm i}$ can
be either larger or smaller than $r_{\pm}$, so this solution applies
to timelike geodesics that originate on either side of $r_\pm$.
We find from \Eq{eq:geodesic} that the geodesic can never cross
the surface at $r=r_{\pm}$: $r(\tau)$ has a turning point when $\tau=-2(r_{\rm i}-r_{\pm})/v_{\rm i}$,
at $r(\tau)=r_{\pm}$. That is, the timelike geodesic can just touch
the surface and bounce off of it, but cannot pass through; see \Fig{fig:bounce}.
Replacing $\tau$ by an arbitrary parameter $\lambda$ respecting the affine connection, the analogous conclusion applies for both null and spacelike geodesics, which also bounce off of this boundary.
(Given $r_{\rm i}$ and $v_{\rm i}$, whether the geodesic  is timelike, null, or spacelike can be specified by choosing the initial data in the $t$ and $\phi$ components of the geodesic equation.)
This justifies
our considering $r_{\pm}$ to be boundaries of the spacetime. 
That is, the connected regions of fixed sign of $N(r)$ (i.e., the three regions $r<r_+$, $r_+ < r < r_-$, and $r>r_-$ in Case II, the two regions $r<r_+$ and $r>r_+$ in Case III, and the two regions $r<r_-$ and $r>r_-$ in Case IV) can be essentially regarded as separate spacetimes.
Indeed, if we replace the right-hand side of the geodesic equation by the
proper acceleration, we see that any trajectory with nonzero ${\rm d}r/d{\rm \tau}$
at $r=r_{\pm}$ will necessarily undergo infinite proper acceleration;
the barrier at $r=r_{\pm}$ is insurmountable.

This example suggests that the spacetime in the region where $N(r)$ is positive, bounded by $r_\pm$, is in
fact geodesically complete. Indeed, we can verify this with the help
of the theorem proven in \Ref{1201.2087}. In the classification
of \Ref{1201.2087}, the family of metrics given in \Eq{eq:metric}
fits the definition of a G\"odel-type spacetime provided $w^{2}(r)+4D(r)>0$.\footnote{Note that the conclusion of \Ref{Bampi}, which exhibited a class of metrics equivalent to the G\"odel universe, does not apply to our metric \eqref{eq:metric}, since we will see in \Sec{sec:Godel} that our class of metrics is strictly larger than the G\"odel solution.}
By \Eq{eq:useful}, this is true whenever $j_{1}\neq0$ and $N(r)>0$.
Hence, our geometry is within the class of
G\"odel-type metrics (a class that includes the G\"odel universe,
some Kerr-Schild metrics, some plane wave metrics, etc.) whenever
we are outside the $N(r)=0$ boundary. Applying Theorem 4.1 of \Ref{1201.2087}, we
observe that since $\left[1-D(r)+\sqrt{(1+D(r))^{2}+w^{2}(r)}\right]^{-1}$
is bounded from above, it follows that our geometry, with $r_\pm$ as its boundary, is geodesically
complete.

\subsection{Censoring CTCs\label{sec:censorship}}

In order to avoid $r_{{\rm C}}>r_{-}$ (while keeping $c_{1}\leq0$
and $c_{2}\geq0$),
we must take either Case II or IV above and further require the tuning
of $c_{1}$ and $c_{2}$ such that
\be
2\frac{c_{1}}{j_{1}}+j_{2}M=\frac{c_{2}}{j_{2}}\label{eq:tuning}
\ee
and also $j_{1}j_{2}<0$. With the choice \eqref{eq:tuning}, the radius of the
outer of the $N(r)=0$ surfaces, which we will call $r_{{\rm H}}$, is located at
\be
r_{{\rm H}}=r_{-}=r_{{\rm C}}=r_0=-\frac{j_{1}}{j_{2}}L=\left|\frac{j_{1}}{j_{2}}\right|L.
\ee
In this case, the radius $r_0$ where the frame-dragging reverses direction, the $N(r) = 0$ surface at $r_-$, and the Cauchy horizon at $r_{\rm C}$ all coincide.
While requiring $c_1 \leq 0$ and $c_2 \geq 0$, it is not possible for the Cauchy horizon to be located a finite distance inside the $N(r)=0$ surface, i.e., $r_{\rm C} \not < r_\pm$.

The choice of coefficients in \Eq{eq:tuning}, necessary to censor
the CTCs, immediately leads us to an observation about the boundary.
Since \Eq{eq:tuning} implies that $D(r_{{\rm H}})=0$, one finds
from \Eq{eq:thetas} that $\lim_{r\rightarrow r_{{\rm H}}}\theta_{k}$
is not zero, but is instead infinite. However, the twist $\omega_{a}$
vanishes as $r\rightarrow r_{\rm H}$ with the choice in \Eq{eq:tuning}.
Hence, for generic $c_{1}$ and $M$, the choice in \Eq{eq:tuning}
is incompatible with a smooth, marginally-trapped event horizon.
We stress that this is not a curvature singularity (neither s.p. nor p.p. type in the sense of \Ref{HawEll}), 
since all curvature scalars for this metric are regular and, moreover, explicit computation shows that the full Riemann tensor is everywhere finite in a local Lorentz frame.\footnote{That is, $e^\mu_{\;\;a} e^\nu_{\;\;b} e^\rho_{\;\;c} e^\sigma_{\;\;d} R_{\mu\nu\rho\sigma}$ is finite, where the dreibein $e^\mu_{\;\;a}$ is defined as $g^{\mu\nu} = e^\mu_{\;\;a} e^\nu_{\;\;b} \eta^{ab}$ for flat metric $\eta$.}
Instead, $r_{\rm H}$ is a boundary in the Lorentzian structure of the manifold, a singularity in the causal structure in the sense of \Ref{gr-qc/9506079}.

\section{Limiting Geometry}\label{sec:limits}

In this section, we will keep the choice in \Eq{eq:tuning} that we made
in \Sec{sec:censorship}, allowing us to exclude the CTCs from
the geometry; as we saw in \Sec{sec:completeness}, we can drop
the region $r<r_{{\rm H}}$ from the geometry with impunity. We would
like to investigate the behavior of the metric for $r\rightarrow\infty$
and for $r\rightarrow r_{{\rm H}}$.

\subsection{Asymptotic geometry\label{sec:asymptotic}}

The metric \eqref{eq:metric} at large $r$ looks like
\be
{\rm d}s^{2}=-{\rm d}t^{2}-2d_{1}\,{\rm d}t\,{\rm d}\phi+d_{2}^{2}\,{\rm d}\phi^{2}+\frac{d_{3}^{4}}{r^{4}}{\rm d}r^{2},
\ee
where $d_{1}=Lj_{2}$, $d_{2}^{2}=c_{2}L^{2}$, and $d_{3}^{4}=\frac{L^{4}j_{1}^{2}}{4\left(j_{2}^{2}+c_{2}\right)}$.
Defining $r'=d_{3}^{2}/r$, $t'=t+d_{1}\phi$, and $d_{4}^{2}=d_{1}^{2}+d_{2}^{2}$,
we have 
\be
{\rm d}s^{2}=-({\rm d}t')^{2}+d_{4}^{2}\,{\rm d}\phi^{2}+({\rm d}r')^{2},
\ee
which is just the geometry of a flat Lorentzian cylinder, $\mathbb{R}^{1,1}\otimes S^{1}$,
of radius $d_{4}$. In particular, at constant $\phi$, this is simply
two-dimensional Minkowski space, and the asymptotic causal
structure includes null infinity. However, the time coordinate $t'$
has a ``jump'' or ``winding'' property due to its dependence on
$\phi$, as discussed in \Ref{Deser:1983tn}.

\subsection{Near-boundary geometry}

Let us define a near-boundary coordinate $x$, where $x^{2}=\frac{r-r_{{\rm H}}}{r_{{\rm H}}}$.
For small $x$, the metric becomes
\be
\begin{aligned}{\rm d}s^{2} & =-{\rm d}t^{2}-2j_{2}Lx^{2}\,{\rm d}t\,{\rm d}\phi-\frac{2L^{3}}{r_{{\rm H}}}\left(c_{1}+j_{1}j_{2}M\right)x^{2}\,{\rm d}\phi^{2}-\frac{j_{2}^{2}Lr_{{\rm H}}{\rm d}x^{2}}{2\left(c_{1}+j_{1}j_{2}M\right)}\\
 & =-{\rm d}t^{2}-2\ell_{3}x^{2}\,{\rm d}t\,{\rm d}\phi+\ell_{1}^{2}x^{2}\,{\rm d}\phi^{2}+\ell_{2}^{2}{\rm d}x^{2},
\end{aligned}
\ee
where we eliminated $c_{2}$ using \Eq{eq:tuning} and defined
$\ell_{1}^{2}=-\frac{2L^{3}}{r_{{\rm H}}}\left(c_{1}+j_{1}j_{2}M\right)$,
$\ell_{2}^{2}=-\frac{j_{2}^{2}Lr_{{\rm H}}}{2\left(c_{1}+j_{1}j_{2}M\right)}$,
and $\ell_{3}=j_{2}L$. Recalling that $c_{1}\leq0$, $j_{1}j_{2}<0$,
and $M\geq0$, we have $c_{1}+j_{1}j_{2}M\leq0$, and hence the ${\rm d}\phi^{2}$
and ${\rm d}x^{2}$ terms are spacelike. The ${\rm d}t\,{\rm d}\phi$
term still induces frame-dragging, but let us consider spatial slices
at constant $t$. The spatial metric is just
\be
\frac{\ell_{1}^{2}}{\ell_{2}^{2}}(x')^{2}{\rm d}\phi^{2}+({\rm d}x')^{2},
\ee
where we have defined $x'=\ell_{2}x$. This is simply the metric of a two-dimensional
plane with a conical defect. Defining $\phi'=\frac{\ell_{1}}{\ell_{2}}\phi$
so that the spatial metric is simply $(x')^{2}({\rm d}\phi')^{2}+({\rm d}x')^{2}$,
$\phi'$ takes the range $0\leq\phi'<2\pi\delta$, where
\be
\delta=\frac{\ell_{1}}{\ell_{2}}=2\left|\frac{c_{1}}{j_{1}}+j_{2}M\right|.
\ee
We could make this conical defect vanish by enforcing $\ell_{1}=\ell_{2}$,
i.e., $\delta=1$, in which case we could simply take the entire surface
at $r=r_{\rm H}$ to be a single point, with a flat spatial metric, albeit with frame-dragging present. Without
this condition, we have a conical singularity if we take $r=r_{\rm H}$
to be identified as a single point. Interpreting the conical defect
as a mass $m$ in a flat background \cite{Deser:1983tn,1610.06101},
we have $m=(1-\delta)/4G$. Since $\delta$ is manifestly positive,
we have $m<1/4G$, so the universe is not overclosed. When $\delta>1$,
this effective mass is negative. Indeed, $m$ is unbounded from
below; as we increase $M$, $\delta$ can grow without limit,
and $m$ can become arbitrarily large and negative.

\section{Symmetries\label{sec:isometries}}

Let us now consider the symmetries of our metric in \Eq{eq:metric}.
For a region where $N(r)>0$ (e.g., $r>\max r_{\pm}$), the
family of geometries described by \Eq{eq:metric} has a Lie algebra
of Killing vectors $v$ each satisfying the Killing equation $\nabla_{(a}v_{b)}=0$. 
The basis of this algebra is given by:
\be
\begin{aligned}u^{a} & =(1,0,0) = \partial_t \\
\phi^{a} & =(0,0,1) = \partial_\phi \\
\chi^{a} & =\sqrt{N(r)}\left(-\frac{2w^{3}(r)\frac{{\rm d}}{{\rm d}r}\left(\frac{D(r)}{w^{2}(r)}\right)}{4D(r)+w^{2}(r)}\sin\alpha\phi,-2\alpha\cos\alpha\phi,\frac{\frac{{\rm d}}{{\rm d}r}\left[4D(r)+w^{2}(r)\right]}{4D(r)+w^{2}(r)}\sin\alpha\phi\right)\\
\psi^{a} & =\sqrt{N(r)}\left(\frac{2w^{3}(r)\frac{{\rm d}}{{\rm d}r}\left(\frac{D(r)}{w^{2}(r)}\right)}{4D(r)+w^{2}(r)}\cos\alpha\phi,-2\alpha\sin\alpha\phi,-\frac{\frac{{\rm d}}{{\rm d}r}\left[4D(r)+w^{2}(r)\right]}{4D(r)+w^{2}(r)}\cos\alpha\phi\right),
\end{aligned}
\ee
where
\be
\alpha=-\frac{2}{j_{1}}\sqrt{2j_{1}j_{2}c_{1}+c_{1}^{2}+j_{1}^{2}\left[\left(M-1\right)c_{2}+j_{2}^{2}M\right]}.\label{eq:alpha}
\ee
The vector $u^{a}$ is the global timelike Killing field we encountered
previously, while the vector $\phi^{a}$ is the angular Killing field
associated with the axial symmetry of the spacetime. The Killing
vectors $\chi$ and $\psi$ exist only if $\alpha$ is an integer, so that $\chi$ and $\psi$ are single-valued. 
That is, for general values of the parameters $(M,j_1 j_2,c_1,c_2)$,
the symmetry algebra is two-dimensional. 

If $\psi$ and $\chi$ are well-defined periodic
Killing vectors, so that the Lie algebra of symmetries is four-dimensional,
then $\alpha = n$, i.e.,
\be
2j_{1}j_{2}c_{1}+c_{1}^{2}+j_{1}^{2}\left[\left(M-1\right)c_{2}+j_{2}^{2}M\right]=\frac{n^{2}j_{1}^{2}}{4}\label{eq:tuning2}
\ee
for some integer $n$. In that case, we can reach any spacetime point
in the connected region where $N(r)>0$ using these Killing vectors, by
moving in $t$ using $u^{a}$, in $\phi$ using $\phi^{a}$, and in
$r$ using $\chi^{a}$ or $\psi^{a}$ in combination with $u^{a}$
and $\phi^{a}$. Hence, as in the G\"odel metric, the isometry group acts transitively and the spacetime region
for $N(r)>0$ is homogeneous.
Thus, if \Eq{eq:tuning2} is satisfied but not \Eq{eq:tuning},
so that CTCs extend to the region where $N(r)$ is positive, the spacetime
is everywhere vicious outside the boundary, with CTCs through every point (so the Cauchy horizon disappears). We will discuss the
relationship between \Eq{eq:metric} and the G\"odel metric further
in \Sec{sec:Godel}. 

\section{Special Cases}\label{sec:special}

Let us now compute some special cases and limits of \Eq{eq:metric}
that are of physical interest, including the choice of parameters
that leads to the G\"odel metric and the behavior of the geometry
in the $\Lambda\rightarrow0$, non-spinning, and $M\rightarrow0$
limits.

\subsection{Relationship to G\"odel universe\label{sec:Godel}}

Choosing 
\be
\begin{aligned}M & =\frac{1}{2}\\
j_{1} & =-2\\
j_{2} & =0\\
c_{1} & =1\\
c_{2} & =0
\end{aligned}
\label{eq:Godelchoice}
\ee
and defining new unitless radial and time coordinates $r'={\rm arcsinh}\sqrt{L/r}$
and $t'=t/\sqrt{2}L$, our metric in \Eq{eq:metric} reduces to
the metric of the G\"odel universe in $2+1$ dimensions \cite{Godel,HawEll}:
\be
{\rm d}s^{2}=2L^2\left[-{\rm d}t'^{2}+{\rm d}r'^{2}-\left(\sinh^{4}r'-\sinh^{2}r'\right){\rm d}\phi^{2}+2\sqrt{2}\sinh^{2}r'\,{\rm d}\phi\,{\rm d}t'\right].
\ee
Hence, our metric constitutes a generalization of the G\"odel metric.
Unlike the G\"odel universe, the general class of metrics in \Eq{eq:metric}
$i)$ corresponds to a matter density $\propto M$ that is a free
parameter, rather than being pinned to the cosmological constant,
$ii)$ has two distinct angular momentum parameters $j_1$ and $j_2$, allowing the
spin of the dust to be nonuniform and even reverse direction at different
radii, and $iii)$ can have apparent horizons for certain choices
of the parameters. 

One can alternatively view the G\"odel solution not as simply describing a dust with negative cosmological constant, but instead as describing arbitrary matter content for which the Einstein tensor in a local Lorentz frame satisfies $G_{ab} \propto {\rm diag}(1,1,1)$.
For example, describing a perfect fluid with arbitrary pressure, plus a cosmological constant, in order to correspond to the G\"odel solution one must satisfy (in $8\pi G = 1$ units) $\rho - p = -2\Lambda$.
That is, in the $\Lambda = 0$ case, the G\"odel universe corresponds to a stiff fluid with $\rho = p$ \cite{Griffiths:2009dfa}.
Similarly, one can view the energy-momentum source for our metric in \Eq{eq:metric} as describing an arbitrary perfect fluid plus cosmological constant for which the Einstein tensor is $L^2 G_{ab} = {\rm diag}(4M-1,1,1)$ in a local Lorentz frame, that is, $\rho + \Lambda = \frac{4M-1}{L^2}$ and $p - \Lambda = \frac{1}{L^2}$.
Taking the cosmological constant to vanish, the equation-of-state parameter $p/\rho = \frac{1}{4M-1}$ corresponds to a stiff fluid when $M = \frac{1}{2}$ in accordance with \Eq{eq:Godelchoice}, a cosmological constant when $M = 0$, phantom energy when $0<M<\frac{1}{4}$, radiation when $M = \frac{3}{4}$, and dust when $M\rightarrow \infty$.

It will be illuminating to impose various energy conditions on this combined perfect fluid \cite{Maeda:2018hqu}.
Imposing the  null energy condition, $R_{ab} k^a k^b \geq 0$ for all null $k^a$, we require $\rho + p \geq 0$, or equivalently, $M \geq 0$. 
Similarly, the strong energy condition, $R_{ab} t^a t^b \geq 0$ for all timelike $t^a$, implies $(D-3)\rho + (D-1)p \geq 0$ in $D$ spacetime dimensions as well as $\rho + p \geq 0$, so in our case it merely stipulates that $M \geq 0$.
Imposing the weak energy condition, $G_{ab} t^a t^b \geq 0$ for all timelike $t^a$, we require $\rho \geq 0$ and $\rho + p \geq 0$, or equivalently, $M \geq \frac{1}{4}$. 
Finally, imposing the dominant energy condition that $-G^a_{\;\;b}t^b$ be causal and future-directed for all causal, future-directed $t^a$, we have $\rho \geq |p|$, so $M \geq \frac{1}{2}$.

Since \Eq{eq:Godelchoice} does not satisfy
\Eq{eq:tuning}, the G\"odel metric exhibits the CTCs for which
it is known.
While the G\"odel universe is in fact totally vicious\textemdash i.e., it
has CTCs through every point, since all points are equivalent as a
consequence of its larger algebra of Killing vectors\textemdash this
is not the case in general. That is, since for general choices of
coefficients in our metric \eqref{eq:metric} the tuning in \Eq{eq:tuning2}
is not satisfied, homogeneity is broken and we have no reason to expect
CTCs for $r>r_{{\rm C}}$. Moreover, if we impose the tuning \eqref{eq:tuning2},
so that the metric for $r>\max r_{\pm}$ is homogeneous, we can sequester
the CTCs behind the boundary if we simultaneously impose \Eq{eq:tuning}.
In that case, homogeneity guarantees that there are no CTCs outside
the boundary. The requirements of Eqs.~\eqref{eq:tuning2} and \eqref{eq:tuning}
can be simultaneously satisfied by first imposing \Eq{eq:tuning}
and then requiring $2|c_1 + j_1 j_2 M| = -j_1 n$.

With the choice of parameters \eqref{eq:Godelchoice}, the parameter
$\alpha$ in \Eq{eq:alpha} is simply $1$, so the G\"odel metric
has four Killing vectors as expected, with \Eq{eq:tuning2} satisfied
with $n=1$. Note that the converse is not necessarily true, however:
satisfying $\alpha=1$ does not imply that the metric is diffeomorphic
to the G\"odel universe, since \Eq{eq:tuning2} does not pin the
dust density to the cosmological constant (i.e., set $M$ to $1/2$)
as is the case in the G\"odel metric.

\subsection{$\Lambda\rightarrow0$ limit}\label{sec:Lambdazero}

For fixed $(M,j_1,j_2,c_1,c_2)$, the $\Lambda\rightarrow0$
(i.e., $L\rightarrow\infty$) limit of the metric described in Eqs.~\eqref{eq:metric}
and \eqref{eq:funs} is singular. However, let us define the rescaled
parameters
\be
\begin{aligned}\iota_{1} & =L^{3}j_{1}\\
\iota_{2} & =Lj_{2}\\
\kappa_{1} & =L^{3}c_{1}\\
\kappa_{2} & =L^{2}c_{2}\\
\mu & =\frac{M}{L^{2}}
\end{aligned}\label{eq:rescaled}
\ee
and then take the $L\rightarrow\infty$ limit, holding $\iota_{1}$,
$\iota_{2}$, $\kappa_{1}$, $\kappa_{2}$, and $\mu$ constant. In this
case, the metric becomes
\be
{\rm d}s^{2}=-{\rm d}t^{2}-2\iota_{2}\,{\rm d}t\,{\rm d}\phi+\Delta(r)\,{\rm d}\phi^{2}+\frac{\iota_{1}^{2}}{4r^{4}\left(\Delta(r)+\iota_{2}^{2}\right)}\,{\rm d}r^{2},\label{eq:flatmetric}
\ee
where
\be
\Delta(r)=-\frac{\iota_{1}^{2}\mu}{r^{2}}+\frac{2\kappa_{1}}{r}+\kappa_{2}.
\ee
This metric satisfies Einstein's equation with zero cosmological constant, for a dust with
uniform density $4\mu$ in the corotating frame,
\be
R_{ab}-\frac{1}{2}R\,g_{ab}=4\mu\,u_{a}u_{b}.
\ee

Defining the winding time coordinate $\bar t=t+\iota_{2}\phi$ as in
\Sec{sec:asymptotic} \cite{Deser:1983tn}, a
new periodic coordinate $\varphi=\phi\sqrt{\kappa_{2}+\iota_{2}^{2}}$,
and a radial coordinate $\rho=|\iota_{1}|/2r\sqrt{\kappa_{2}+\iota_{2}^{2}}$,
the metric \eqref{eq:flatmetric} becomes

\be
{\rm d}s^{2}=-\left({\rm d}\bar t\right)^{2}+f(\rho)\,{\rm d}\varphi^{2}+\frac{{\rm d}\rho^{2}}{f(\rho)},\label{eq:metdS}
\ee
where
\be
f(\rho)=1-4\mu\rho^{2}+\frac{4\kappa_{1}}{|\iota_{1}|\sqrt{\kappa_{2}+\iota_{2}^{2}}}\rho.\label{eq:fdS}
\ee
We note, remarkably, that if $\kappa_{1}=0$, then the spatial sector
of \Eq{eq:metdS} is simply a Euclideanized two-dimensional de
Sitter space in static slicing; to make the analogy precise, one would
require $\kappa_{2}+\iota_{2}^{2}=1/4\mu$, so that the periodicity
in $\text{\ensuremath{\varphi}}$ (the Wick-rotated time coordinate
of the $\text{dS}_{2}$ space) matches the dS length.

\subsection{Non-spinning limit}

The geometry described in Eqs.~\eqref{eq:metric} and \eqref{eq:funs}
must always be spinning; there is no way to smoothly take the joint
$j_{1},j_{2}\rightarrow0$ limit while keeping the metric nonsingular.
Specifically, the form of $N(r)$ and $D(r)$ dictate that we cannot 
take $j_{1}\rightarrow0$. 

However, we can set $j_{2}$ to zero. In this case, the ${\rm d}t\,{\rm d}\phi$
term in the metric asymptotes to zero for large $r$. The vorticity tensor
\eqref{eq:twisttensor} for the timelike congruence along $\partial_{t}$
then satisfies $\lim_{r\rightarrow\infty}\Omega_{ab}=0$, but the
twist one-form for the null geodesics described in \Sec{sec:apparent}
satisfies $\lim_{r\rightarrow\infty}\omega_{a}=\left(0,0,\mp\sqrt{c_{2}}\right)$,
where the $+$ case occurs if and only if $j_{1}$ is negative (and vice versa). Thus, even when we
send $j_{2}$ to zero, information about the spin of the dust is imprinted
on the twist of null geodesics at arbitrarily large $r$, even though the frame-dragging $w(r)$ asymptotically vanishes.

\subsection{Vacuum limit\label{sec:vacuum}}

Let us consider the vacuum limit of the geometry described in \Eq{eq:metric}.
Taking the gas density ($\propto M$) to zero, we find that the Cauchy horizon is located at
\be
r_{{\rm C}}=-2\frac{c_{1}}{c_{2}}L.
\ee
For $r>r_{{\rm C}}$, let us define
\be
\tilde{r}=L\sqrt{\frac{2c_{1}L}{r}+c_{2}},\label{eq:rredefine}
\ee
so $D(r){\rm d}\phi^{2}\rightarrow\tilde{r}^{2}{\rm d}\phi^{2}$.
If we continue to impose the condition \eqref{eq:tuning} (which has simply become $c_2/j_2 = 2c_1 /j_1$ for $M=0$) in order to sequester
the CTCs behind a boundary, then the metric becomes
\be
{\rm d}s^{2}=-{\rm d}t^{2}-\frac{j_{1}\tilde{r}^{2}}{c_{1}L}{\rm d}t\,{\rm d}\phi+\tilde{r}^{2}{\rm d}\phi^{2}+\left(\frac{4c_{1}^{2}}{j_{1}^{2}}+\frac{\tilde{r}^{2}}{L^{2}}\right)^{-1}{\rm d}\tilde{r}^{2}.
\ee
The boundary is located at $\tilde{r}=0$, so the CTC region behind
the boundary corresponds to the analytic continuation of $\tilde{r}^{2}$
to negative values, cf. \Ref{gr-qc/9506079}.

Next, we can rescale the radial coordinate again, defining $\hat{r}=\left|\frac{j_{1}}{2c_{1}}\right|\tilde{r}$,
and also define $\hat{\phi}=\frac{2c_{1}}{j_{1}}\phi$, $0\leq|\hat{\phi}|<\left|\frac{4\pi c_{1}}{j_{1}}\right|$,
so
\be
{\rm d}s^{2}=-{\rm d}t^{2}-\frac{2\hat{r}^{2}}{L}{\rm d}t\,{\rm d}\hat{\phi}+\hat{r}^{2}\,{\rm d}\hat{\phi}^{2}+\left(1+\frac{\hat{r}^{2}}{L^{2}}\right)^{-1}{\rm d}\hat{r}^{2}.\label{eq:intermediate}
\ee
We can turn \Eq{eq:intermediate} into a metric that locally corresponds
to $\text{AdS}_{3}$ in global coordinates by defining $\bar{\phi}=\hat{\phi}-\frac{t}{L}$, in terms of which we have
\be
{\rm d}s^{2}=-\left(1+\frac{\hat{r}^{2}}{L^{2}}\right){\rm d}t^{2}+\left(1+\frac{\hat{r}^{2}}{L^{2}}\right)^{-1}{\rm d}\hat{r}^{2}+\hat{r}^{2}\,{\rm d}\bar{\phi}^{2}.\label{eq:AdS2}
\ee
If $t$ is allowed to take values in all of $\mathbb{R}$, then
$\bar{\phi}$ spans the real numbers and the metric in \Eq{eq:AdS2} describes
a covering space of ${\rm AdS}_{3}$ with decompactified time and
angular coordinate. To correspond to $\text{AdS}_{3}$, we should
make $t$ periodic, $0\leq t<2\pi L$. For consistency, we can then
take $2c_{1}/j_{1}=1$ so that $\hat{\phi}\in[0,2\pi)$, and we are
left with the global ${\rm AdS}_{3}$ geometry. We note that if $2c_{1}/j_{1}$
is any integer $n$, we have $\hat{\phi}\in[0,2\pi n)$, which (for
appropriate periodicity in $t$) corresponds simply to an $n$-fold
cover of ${\rm AdS}_{3}$, which we may quotient by $\mathbb{Z}_{n}$
to recover the single copy of the geometry. For any $n$, \Eq{eq:tuning}
and our $M=0$ choice together then imply that \Eq{eq:tuning2} is satisfied, enhancing the symmetry algebra (which is expected, since ${\rm AdS}_3$ in fact possesses six Killing vectors). 

Hence, we find that in the CTC-free case where we impose \Eq{eq:tuning},
our metric \eqref{eq:metric} obeys the extension of Birkhoff's theorem
to $2+1$ dimensions \cite{0403227}, which states that any solution
of the $(2+1)$-dimensional vacuum Einstein equations with negative
cosmological constant that is free of CTCs must correspond to ${\rm AdS}_{3}$,
a BTZ geometry \cite{BTZ}, or a Coussaert-Henneaux \cite{Coussaert} solution.

\section{Discussion}\label{sec:discussion}

The metric in \Eq{eq:metric} contains six free parameters: a length $L$ and five unitless constants $(M, j_1, j_2, c_1, c_2)$. Choosing a value of $L$ effectively sets the length scale of the geometry (relative to the three-dimensional Newton's constant). The remaining five parameters specify a five-dimensional space of geometries that, for general choices of the constants, remain distinct under diffeomorphisms. 
The parameter $M$ can be measured in the geometry through the value of $R_{tt}$, while $c_2$ can be measured by the proper circumference of the spacetime at large $r$. 
Moreover, the existence and location of the zeros of $N(r)$, $D(r)$, and $w(r)$ (i.e., $r_\pm$, $r_{\rm C}$, and $r_0$, respectively) allow various combinations of the remaining three parameters to be measured in the spacetime.
These locations are all physically meaningful; the $rr$ component of the metric flips sign at $r_\pm$, CTCs appear at $r_{\rm C}$, and the frame-dragging vanishes at $r_0$. The relative order and proper distances among these locations and the axis of rotation allow us to conclude that $(M,j_1,j_2,c_1,c_2)$ indeed specifies a five-dimensional family of globally inequivalent geometries.

This paper leaves multiple interesting avenues for further research.
We leave for future work the subjects of the dynamical stability of these geometries and their formation from the collapse of a cloud of dust.
The question of what other three-dimensional geometries exhibit an analogue of the bouncing boundary discussed in \Sec{sec:completeness} is also a compelling one.
Further, it would be interesting to study the causal structure of the $c_1 > 0$ case mentioned in \Sec{sec:apparent}, in which an apparent horizon appears at radius $r_{\rm m}$ and $D(r)$ asymptotically decreases.
The appearance of the analytically-continued de Sitter metric in \Sec{sec:Lambdazero} also merits investigation; the double Wick rotation is reminiscent of the ``bubble of nothing'' solution of \Ref{Witten:1981gj}.
Finally, an examination of the spatial geodesics of this class of metrics, how they can be embedded within a surrounding vacuum to produce an asymptotically-AdS geometry, and a subsequent computation of their Ryu-Takayanagi surfaces~\cite{Ryu:2006bv} are well-motivated future directions from a holographic perspective.
 
\begin{center} 
{\bf Acknowledgments}
\end{center}
\noindent 
It is a pleasure to thank Raphael Bousso, David Chow, Metin G\"urses, Illan Halpern, and \linebreak Yasunori Nomura for useful discussions and comments.  
This work is supported by the Miller Institute for Basic Research in Science at the University of California, Berkeley.

\appendix
\section{Gaussian Dust Solution}\label{app:Gaussian}

In this appendix, we discuss an additional, apparently new solution of the Einstein equations in $2+1$ dimensions. Although not globally a member of the family of metrics presented in \Eqs{eq:metric}{eq:funs}, we will find that it is related in a certain limit. Consider a dust solution to the Einstein equations with zero cosmological constant,
\be
R_{ab} - \frac{1}{2} R\,g_{ab} = m(r) \,u_a u_b,
\ee
with vector $u^a = \partial_t$ as before. Here, let us take the density $m(r)$ to have a Gaussian profile,
\be 
m(r) = \frac{1}{a^2} e^{-r^2/a^2},
\ee
for some length parameter $a$. The metric for the solution is
\be
{\rm d}s^2 = -{\rm d}t^2  - 2a\,{\rm d}t\,{\rm d}\phi +(r^2 - a^2){\rm d}\phi^2 + e^{r^2/a^2} {\rm d}r^2.\label{eq:gaussianmetric}
\ee
The solution has a two-dimensional isometry algebra generated by Killing vectors $u^a$ and $\phi^a$. Since $u^a \nabla_a u^b = 0$, $u^a$ generates a timelike congruence as before, for which we find that the expansion $\theta$, shear $\varsigma$, and vorticity $\Omega$ all vanish. The solution is clearly distinct from G\"odel's universe \cite{Godel} (since the dust density is not uniform) and van~Stockum's solution \cite{VanStockum} (since the dust density is finite in the closure of the geometry and goes to zero at the boundary). Such a rotating dust solution, with higher density in the center and exponentially-falling density at large distances, may be of astrophysical use if extended to four dimensions. In a sense, this solution may be thought of as corresponding to a $(2+1)$-dimensional galaxy. However, the solution possesses CTCs for $r<a$.

Taking the small-$r$ limit of the $rr$ component of the metric in \Eq{eq:gaussianmetric}, we can write, for $r\ll a$,
\be
{\rm d}s^2 \simeq -{\rm d} \hat t^2 + r^2 {\rm d}\phi^2  + \frac{{\rm d}r^2}{1-\frac{r^2}{a^2}},\label{eq:approxmetric}
\ee
where we define $\hat t = t+a \phi$.
The metric \eqref{eq:approxmetric} is a member of the family of $\Lambda = 0$ solutions discussed in \Sec{sec:Lambdazero}; if we set the parameters in \Eq{eq:rescaled} to $\mu = 1/4a^2$, $\kappa_1 =\iota_2 = 0$, $\kappa_2 = a^2$, write $t$ in \Eq{eq:flatmetric} as $\hat t$, and send $r$ in \Eq{eq:flatmetric} to $|\iota_1|/2a\sqrt{a^2 - r^2}$, we recover \Eq{eq:approxmetric}. Thus, as we would expect, the small-$r$ (and thus nearly-constant density) limit of the metric \eqref{eq:gaussianmetric} corresponds to a member of the family of uniform-density rotating dust solutions in \Eq{eq:metric}.

\noindent

\bibliographystyle{utphys-modified}
\bibliography{rotating_dust}

\end{document}